      \documentstyle[preprint,prc,aps,floats,epsf]{revtex}
\begin{document}

% uncomment \draft to have PACS numbers appear
\draft

% put preprint number(s).
\preprint{\vbox{\hfill NUC--MINN--96/11--T}}

\title{A New Approach to Chiral Perturbation Theory \\
           with Matter Fields}

\author{Hua-Bin Tang}
\address{ School of Physics and Astronomy\\
          University of Minnesota, Minneapolis, MN\ \ 55455}

%\date{\today}
\date{July 9, 1996}
\maketitle

\begin{abstract}
I propose a prescription for separating the high- and low-energy 
contributions in effective field theories. This prescription 
allows a relativistic treatment of matter fields in chiral 
perturbation theory while the power counting remains valid.
\end{abstract}


\narrowtext

Quantum chromodynamics (QCD) is the underlying theory of low-energy
effective field theories (EFTs). Despite the presence of an infinite 
number of free parameters, an EFT may provide
useful perturbative calculations 
in powers of the energy\cite{WEINBERG79,GASSER84}.
Underlying the low-energy expansion is the power counting of
Weinberg\cite{WEINBERG79,WEINBERG90}.

In general, EFTs permit useful low-energy expansions
only if we absorb {\it all} the
hard-momentum effects into the parameters of the
lagrangian. For chiral perturbation theory (CHPT)
in the pion sector\cite{GASSER84}, 
the only hard-momentum
effects are the ultraviolet divergences, which are absorbed into 
the parameters through renormalization. 
When  we include the nucleons relativistically,
the anti-nucleon  contributions are also  hard-momentum
effects. These hard effects were not absorbed into the parameters of
the lagrangian in the relativistic treatment of 
Gasser, Sainio, and Svarc (GSS)\cite{GASSER88}. As a result, 
the power counting fails and a systematic
expansion in energy is not possible.
In the heavy-baryon formalism, Jenkins and Manohar\cite{JENKINS91}
recover the power counting by integrating out the anti-nucleon 
field so that
its effect is  absorbed into the parameters.
Weinberg\cite{WEINBERG90} has also introduced a similar
nonrelativistic formalism for the nucleons in discussing the 
nucleon--nucleon potential.

Although heavy-baryon ChPT has been successful in many applications,
it is useful to have a relativistic formalism in which the power
counting is still valid\cite{LUTZ}. First, we do not need to keep track of 
various sets of terms of $1/M$ corrections
resulting from the nonrelativistic reduction of the 
lagrangian\cite{ECKER96}, 
where $M$ is the nucleon mass. 
Next, a relativistic formalism may provide new insights
into the momentum expansion.  
For example, as demonstrated later, I find  that loops
may generate divergences to all orders without jeopardizing 
the power counting---a claim that may supprise 
some practioners in CHPT. Finally,
a relativistic description  also significantly simplifies
the finite-density problem,  where the $1/M$ corrections 
as well as the differences between the scalar density 
$\langle\overline{N} N\rangle$
and the vector density $\langle N^\dagger N\rangle$ 
are important. Indeed, the relativistic many-body hadronic theory,
so-called quantum hadrodynamics (QHD)\cite{SEROT86}, 
has been quite efficient for calculations in nuclear matter
and finite nuclei.

To achieve a useful relativistic formalism with matter 
fields (i.e. those of heavy hadrons),
we must find some unambiguous prescription to separate the hard- and 
soft-momentum effects so that we can absorb the former
into the parameters while evaluating the latter.
We may consider the absorption of the hard-momentum parts
of any Feynman diagrams into the parameters as an extra renormalization.
This extra renormalization is necessary for a useful low-energy
expansion because we have a theory that
contains not only Goldstone bosons whose masses are much lighter than
the large-energy scale, but also  baryons whose masses
are of the same order as this scale.
Note that the idea of absorbing vacuun effects of heavy hadrons
into the parameters has been adopted in Ref.\cite{FST95}, where
the vacuum one-baryon-loop energy of nuclear matter and finite nuclei
are absorbed into the scalar potential.  
In this Letter, I propose a prescription for 
separating the hard- and soft-momentum
contributions in any Feynman diagrams so that we can perform
the extra renormalization. I show that the soft part
of any Feynman diagram satisfies Weinberg's power counting,
so a systematic low-energy expansion becomes possible even
in the relativistic theory with matter fields.

We start with the same
relativistic chiral lagrangian as in GSS and use the 
equations of motion or field redefinitions to remove redundant
terms\cite{WEINBERG90,GEORGI91,ARZT95}.
The lagrangian without external sources and with the inclusion
of the $\Delta$ isobar can be found in Refs.~\cite{TE96,ET96}.
Thus, we derive the Feynman rules just as usual. 
We do not deviate from GSS until we deal with the loop integrals 
and perform the extra renormalization which absorbs the hard parts
into the parameters of the lagrangian.  GSS keep
the large hard parts that spoil the power counting.
Jenkins and Manohar\cite{JENKINS91} modify the relativistic lagrangian
by integrating out the anti-nucleon field.
In this heavy-baryon formalism, they achieve
a useful low-energy expansion. However,
their lagrangian is much more involved 
than the original relativistic one\cite{ECKER96}. 
Our approach will  produce the same results as the heavy-baryon
formalism if the $1/M$ corrections are included appropriately
in the latter. The advantage of our approach  is that it enjoys the
simplicity of the relativistic lagrangian as in the
treatment of GSS, while it also allows a
consistent low-energy expansion as in the
heavy-baryon formalism.

I emphasize that loops may generate divergences 
to all orders so that we may require renormalizaton to all orders
at any loop order. This requirement does not cause  problems
because the truncation of the lagrangian must be
based on renormalized coefficients; it is these renormalized
coefficients that may satisfy the naive dimensional analysis
\cite{MANOHAR84,GEORGI93b}. We simply  need to retain contact
terms up to the same order as that of the renormalized 
loop contributions, rather than up to 
the highest order in which the loops 
generate divergences.

Let $Q$ stand for a generic
soft-momentum scale of the order of the pion mass $m_\pi$
and take the nucleon mass $M$ to represent 
the typical large-energy scale.
We are interested in applications where
the space components of external
nucleons are of order $Q$ and so is the mass 
splitting $(M_\Delta -M)$
between the nucleon and the $\Delta$-isobar. Quantities of
order $M$ include the $\Delta$ mass and the factor $4\pi f_\pi$
with $4\pi$ coming from a loop integral\cite{MANOHAR84} and
$f_\pi\approx 93\,$MeV being the pion decay constant.

We can write the integral $G$ for a Feynman graph as the sum of
its hard and soft parts:
\begin{equation}
          G = (G- \hat{R}\hat{S}G) + \hat{R}\hat{S}G \ , \label{eq:sep}
\end{equation}
where $\hat{S}$ is an operator that extracts the unrenormalized
``soft'' part of $G$ and $\hat{R}$ is an operator that 
renormalizes $\hat{S} G$
in the standard way so as to remove the residual hard contributions
(ultraviolet divergences in the form of poles at $d=4$ 
in dimensional regularization). 
The prescription for obtaining the unrenormalized
soft part $\hat{S} G$ from G 
consists of the following rules: 
\begin{itemize}
\item[1.] Take the loop momenta to be of order $Q$.
\item[2.] Make a covariant $Q/M$ expansion of the integrand.
\item[3.] Exchange the order of the integration and 
          summation of the resulting power series.
\end{itemize}

That the prescription indeed extracts the soft part
of a Feynman diagram is shown as follows.
Let $q$ be a loop momentum and
consider the integration over the time component $q_0$. Closing
the contour by a semicircle at infinity, we find that the
$q_0$-integration of $G$ is given by
the sum of three contributions: (1) the semicircle, (2) the soft
poles, and (3) the hard poles. Here
the soft and hard poles are those at momenta of orders $Q$ 
and $M$ respectively. We have assumed that 
we can devide the poles into the soft and hard ones,
as is the case for theories with Goldstone bosons and
massive baryons. First, we note that the 
semicircle could produce divergences which
will later be removed by the usual renormalization. Next,
we note that our prescription does not allow
a $Q/M$ expansion of the soft-pole structures such as the pion 
propagator. However, we can make a $Q/M$ expansion of
hard-pole structures because the momenta at which the poles locate 
are much larger than the loop momentum 
when the latter is taken to be of order $Q$ by Rule 1. 
We require a covariant expansion to maintain Lorentz invariance.
Finally, integration term by term by Rule 3 
removes the contributions of the hard poles since they
do not appear in any individual terms. Thus, we are left with
the unrenormalized soft part, in which
ultraviolet divergences can still occur.

Because the prescription retains all the soft poles, the remainder
$(G-\hat{R}\hat{S}G)$ must have contributions only
from large momenta including some ultraviolet divergences.
We can write this hard part as local counterterm contributions---a
well-known result that is fundamental to the idea of effective field
theories. This result is intimately related to the
uncertainty principle as argued by Lepage\cite{LEPAGE89}. Indeed,
large momenta correspond to short distances that are 
tiny compared with the wavelengths of the external particles,
so the interactions must be local. Thus, we can perform
the extra renormalization of absorbing the hard part
into the parameters of the lagrangian.
That Weinberg's power counting remains valid in our 
approach can be shown as follows:

First, we prove 
that diagrams with closed fermion loops contain no soft parts.
Two kinds of fermion loops may occur.
The first  kind does not contain any
nearly on-shell baryon lines. An example is a baryon loop that is not
connected  to any external baryon lines. 
In this case, Rules 1 and 2 allow us to expand the fermion propagators
as polynomials in the loop momentum. For example, we can expand
the nucleon propagator $G(q)$ as follows:
\begin{equation}
  G(q) =         { 1 \over \rlap/{\mkern-1mu}q
                       - M + i \epsilon }
         = -{1\over M}\bigg(1 +{\rlap/{\mkern-1mu}q \over  M}
             + {q^2\over M^2} + \cdots \bigg) 
             \ , \label{eq:expN1}
\end{equation}
where $q$ is taken to be of order $Q$ by Rule 1. 
Thus, integrating term by term using dimensional regularization,
we obtain vanishing soft parts. 
The second kind may involve multiple
fermion loops connected to external and containing 
internal baryon lines that are nearly on shell. This kind has
at least one fermion loop that contains one and only one 
nearly on-shell baryon line.
We note that an internal nearly on-shell baryon propagator
carries a momentum $k+q$ 
with both $q$ and $(k^2-M^2)/M$ being of order $Q$. Thus,
for a nucleon like this, we obtain
the covariant expansion of its propagator as
\begin{equation}
  G(k+q)
        = {\rlap/{\mkern-2mu}k+
               \rlap/{\mkern-1mu}q + M \over
              2k\cdot q+ k^2 - M^2 + i\epsilon}
             -{q^2 (\rlap/{\mkern-2mu}k+
               \rlap/{\mkern-1mu}q + M )
            \over (2k\cdot q+ k^2 - M^2 + i\epsilon)^2}
             +\cdots
                      \ , \label{expN}
\end{equation}
where the leading term is of order $1/Q$
and each succeeding term is suppressed by $Q/M$. 
Now for the loop
with just one nearly on-shell baryon line, by Rules 1 amd 2,
we can  expand
the nearly on-shell propagator
as in Eq.~(\ref{expN}) and all other
baryon propagators as in  Eq.~(\ref{eq:expN1}).
Rule 3 then allows us to integrate term by term.
The resulting loop integral 
again vanishes in dimensional regularization since
(see Ref.~\cite{COLLINS84} for example)
\begin{equation}
     I(k) \equiv      \int\!{\rm d}^d{q}
           {  (q^2)^m
           \over (2k\cdot q+ k^2 - M^2 + i\epsilon)^n } = 0
                           \label{eq:Ik}
\end{equation}
for any integers $m$ and $n$ and any derivatives of $I(k)$ also
vanish. Thus, we can ignore fermion loops.

Next, we note that, once the baryon loops are excluded, 
any baryon lines must be connected to some external baryons
by baryon current conservation. Thus, the baryon propagators are
nearly on shell and are
of order $1/Q$ from the expansion in Eq.~(\ref{expN}).
In practice, we truncate Eq.~(\ref{expN}) at some order
depending on the accuracy to which
we calculate the relevant physical quantity. 
The rest of the power-counting argument becomes the same as 
in Weinberg's work\cite{WEINBERG90}. It follows
that a Feynman diagram with $L$ loops, $E_{\rm N}$ external baryon lines
carries the order $Q^\nu$ with
\begin{equation}
      \nu = 2+2 L -\case{1}{2}E_{\rm N} 
          + \sum_{i} V_i \Big (d_i + \case{1}{2} n_i -2\Big) 
                 \ , \label{eq:chcnt}
\end{equation}
where $V_i$ is the number of
vertices of type $i$ characterized by $n_i$ baryon
fields and $d_i$ pion derivatives or $m_\pi$ factors.

\begin{figure}
 \setlength{\epsfxsize}{2.2in}
  \centerline{\epsffile{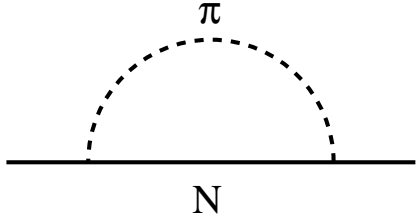}}
%\vspace*{.2in} 
\caption{One-loop self-energy of the nucleon.}
 \label{fig:one}
%\vspace*{.4in}
\end{figure}

We have shown that the prescription can be carried out for any
Feynman diagrams. Thus, it can also be carried out for any
Green's functions because they are just the sum
of Feynman diagrams constructed from the usual Feynman
rules, which are still applicable in our  approach.
Note that the prescription 
has no effects on ChPT in the pion sector because
there are no hard poles there. Also,
in many practical calculations, the hard part is seldom needed
because it is cancelled by counterterm contributions. Instead, we 
evaluate and renormalize the soft part  
and combine it
with the tree contributions, which we truncate up to the same order 
as the loop contributions.

In what follows, to illustrate how the prescription works, 
I calculate in detail
the nucleon one-loop self-energy with a $\pi N$ intermediate
state, as shown in Fig.~\ref{fig:one}. 
Both the hard and soft parts will be evaluated explicitly.
The corresponding result in heavy-baryon ChPT can be 
reproduced in the limit of infinite nucleon mass.
I shall also briefly discuss the structure of the $\pi N$
scattering amplitude up to order $Q^3$.

From the Feynman rules derived as usual, 
the self-energy of a nearly on-shell nucleon of momentum $k$ is
\begin{equation}
\Sigma_{\pi\rm N}(k) = -
             { 3g_{\rm A}^2 \over 4 f_\pi^2 }\, i\mu^{4-d}
         \!   \int\!{{\rm d}^d{q}\over ( 2\pi)^d}
           {  \rlap/{\mkern-1mu}q \gamma_5
              G(k+q)\,
              \rlap/{\mkern-1mu}q \gamma_5
              \over q^2 - m_\pi^2 + i \epsilon }
                      \ , \label{sigpiN}
\end{equation}
where $g_{\rm A}\approx 1.26$ is the axial coupling 
and $\mu$ is the scale of dimensional regularization.
According to
the prescription, we obtain the soft part of $\Sigma_{\pi\rm N}$
by first making a covariant $Q/M$ expansion of the integrand while
taking the loop momentum $q$ to be soft. Since we assume
$k$ to be nearly on shell, the expansion of the nucleon propagator
is as given in Eq.(\ref{expN}).
Exchanging the order of the summation and integration
then gives the soft part.

Although we may truncate
the expansion in Eq.~(\ref{expN}) at some order, 
we notice in the present case
that, after our exchange of the integration and summation
and with the use of Eq.~(\ref{eq:Ik}), 
the $q^2$ in the numerator of any integrand 
can be replaced
with $m_\pi^2$ in dimensional regularization.
After this replacement we may
resum  the series and postpone the discussion of the $1/M$ expansion
until the integral is performed.  Thus, we
obtain the one-loop self-energy to all orders in the $1/M$ expansion:
\begin{equation}
\hat{S} \Sigma_{\pi \rm N}(k) 
           = {3\over 4}
                  { g_{\rm A}^2 \over f_\pi^2 }\, i\mu^{4-d}\!
           \int\!{{\rm d}^d{q}\over ( 2\pi)^d}
           {  (\rlap/{\mkern-2mu} k + M)m_\pi^2
                 - (2 k\cdot q +m_\pi^2)\rlap/{\mkern-1mu}q 
           \over (q^2 - m_\pi^2 + i \epsilon )
                 (2k\cdot q+ k^2 - M^2 + m_\pi^2 + i\epsilon) }
               \ . \label{eq:piN}
\end{equation}
Clearly, the soft part $\hat{S} \Sigma_{\pi \rm N}$ is of order
$Q^3/M^2$ in agreement with Eq.~(\ref{eq:chcnt}). It is
straightforward to perform the integral of Eq.~(\ref{eq:piN})
to yield
\begin{eqnarray}
\hat{S}\Sigma_{\pi\rm N} (k) &=&
        \Sigma^{\overline{\rm MS}}_{\pi\rm N}(k) 
                +{3 g_{\rm A}^2\over 4f_\pi^2}
                \bigg [ {k^2-M^2 \over k^2} (m_\pi^2-2\omega^2)
                 \,  \rlap/{\mkern-2mu}k \nonumber \\[3pt]
         &    & \ \ \ \ \ \ \ \ \ \ \ \ \ \ \ \ \ \
                +{ 2\omega m_\pi^2 \over \sqrt{k^2} }
                      (\rlap/{\mkern-2mu}k + M)\bigg]
               \bigg( L +{1\over 32\pi^2} \ln {m_\pi^2 \over \mu^2} 
                           \bigg)
                                \ , \label{eq:Nsig} \\[3pt]
        \Sigma^{\overline{\rm MS}}_{\pi\rm N}(k) 
               & = &
                {3 g_{\rm A}^2\over 64\pi^2 f_\pi^2}
                \bigg [ {k^2-M^2 \over k^2} \omega
                  \, \rlap/{\mkern-2mu}k 
                -{ m_\pi^2 \over \sqrt{k^2} }
                      (\rlap/{\mkern-2mu}k + M)\bigg]
                        \nonumber \\[3pt]
             & &  
           \times \bigg( \omega + 2  \sqrt{m_\pi^2-\omega^2}
              \, \tan^{-1}{\sqrt{m_\pi^2-\omega^2}\over \omega} \bigg)
                                \ , \label{eq:NMS}
\end{eqnarray}
where we introduce the following notation:
\begin{eqnarray}
     \omega &\equiv&
           {1\over 2\sqrt{k^2} }(k^2-M^2 +m_\pi^2) \ ,
                    \\[3pt]
      L     &\equiv & {1\over 32\pi^2}
            \Big( {2\over d-4}+\gamma -1 -\ln 4\pi \Big) \ ,
\end{eqnarray}
with  Euler's constant $\gamma = 0.577\cdots$.
As seen from Eq.~(\ref{eq:Nsig}), $\Sigma^{\overline{\rm MS}}_{\pi\rm N} $
is the self-energy renormalized
at scale $\mu=m_\pi$
in the modified minimal subtraction 
($\overline{\rm MS}$) scheme.

Expanding the second term of Eq.~(\ref{eq:Nsig}) around 
$\rlap/{\mkern-1mu}k = M$, we find that the divergences appear
up to infinite order in powers of 
$(\rlap/{\mkern-1mu}k-M)/M$. This is the same 
low-energy expansion employed 
for the lagrangian~\cite{GASSER88,TE96,ET96} except that we 
use the equations of motion\cite{WEINBERG90,GEORGI91,ARZT95} 
to remove higher derivatives on the
nucleon fields in favor of multiple $\pi N$ and $NN$ interaction terms.
For example, counterterms such as 
$
{1\over M}\overline{N}
       (\rlap/{\mkern-1mu} \partial - M)^2 N 
$
are reduced to other interaction
terms by the nucleon equation of motion.
Thus, we can absorb all the divergences
into the infinite number of parameters in the lagrangian. 
As emphasized before, this does not invalidate the power counting
even though we need an infinite number of counterterms
for any loops.

Note  further that
$ \Sigma^{\overline{\rm MS}}_{\pi\rm N}$, the self-energy 
in the $\overline{\rm MS}$ scheme at $\mu=m_\pi$, is non-analytic
and so it cannot be absorbed into the parameters of the lagrangian. 
This result is consistent with the expectation that the parameters should
contain only high-energy contributions.
Finally, taking the limit of infinite
nucleon mass, we can verify straightforwardly that
Eq.~(\ref{eq:Nsig}) reduces to the corresponding
result in heavy-baryon ChPT, as in e.g. Ref.~\cite{BERNARD92}.

It is reassuring to verify that the hard part can be absorbed
into the parameters of the lagrangian.
Performing the integral in Eq.~(\ref{sigpiN}) 
directly, we obtain
\begin{eqnarray}
(1-\hat{S})\Sigma_{\pi\rm N} (k) &=&
               {3 g_{\rm A}^2\over 2f_\pi^2} M^2
               \bigg ( M
                    +{k^2+M^2 \over 2k^2}\rlap/{\mkern-2mu}k
                \bigg)
                 \bigg( L +{1\over 32\pi^2}
                          \ln {M^2 \over \mu^2} \bigg)
                  \nonumber \\[3pt]
 &    & 
      - {3 g_{\rm A}^2\over 64\pi^2f_\pi^2} 
           \Big(\sqrt{k^2}-\omega\Big)
               \bigg[
                 {k^2-M^2 \over k^2} \omega
                  \, \rlap/{\mkern-2mu}k
                -{ m_\pi^2 \over \sqrt{k^2} }
                      (\rlap/{\mkern-2mu}k + M)\bigg]
                  \nonumber \\[3pt]
      &    & \ \ \ \  \
               \times  \bigg[ 32\pi^2 L
                       +   \ln {M^2 \over \mu^2} 
             - 1 + \sum_{l=1}^{\infty} {2  \over 2 l -1}\,
             {(\omega^2 - m_\pi^2)^l \over
                      (\sqrt{k^2}-\omega)^{2l} }\, \bigg]
                      \ , \label{eq:hard}
\end{eqnarray}
which is indeed expandable in powers of 
$(\rlap/{\mkern-1mu} k-M)/M$. Notice, however, 
that the first term in
Eq.~(\ref{eq:hard}) is of $O(M)$ and the second one is of $O(Q^2/M)$,
both of which are of higher order than the soft part, 
which is of $O(Q^3/M^2)$. This result is consistent because the
hard part comes from large-momentum contributions that spoil
the power counting. Note that even in the relativistic treatment
of GSS\cite{GASSER88} the first term in
Eq.~(\ref{eq:hard}) with $\rlap/{\mkern-1mu} k = M$ was absorbed
into the bare nucleon mass.

Although the $\overline{\rm MS}$ scheme may be sufficient 
for the nonrenormalizable higher-order terms,
we should perform mass and wave-function renormalizations
on the nucleon mass shell because the nucleons are physically observed. 
Thus, we need further mass and wave-function counterterm
subtractions to obtain the renormalized self-energy:
\begin{equation}
  \hat{R}\hat{S}\Sigma_{\pi\rm N}(k)
          = \Sigma^{\overline{\rm MS}}_{\pi\rm N}(k)
           -\Sigma^{\overline{\rm MS}}_{\pi\rm N}(k)
           \Big |_{\rlap/{\mkern-1mu} k = M}
            -{\partial \over \partial \rlap/{\mkern-1mu} k}
             \Sigma^{\overline{\rm MS}}_{\pi\rm N}(k)
             \Big |_{\rlap/{\mkern-1mu} k = M}
                 (\rlap/{\mkern-2mu} k - M)  
                               \ . \label{eq:NslfE}
\end{equation}
Note that we have adopted the counterterm method of renormalization 
by starting with physical masses and couplings and then
adding counterterms. To unveil the chiral expansion of 
the nucleon mass we can use
\begin{equation}
  M = M_0 + \Sigma_{\pi\rm N}(k) |_{\rlap/{\mkern-1mu} k = M}
                      \ , \label{eq:Nmass}
\end{equation}
where $M_0$ denotes the sum of the bare nucleon mass and the bare
contributions from the symmetry-breaking contact terms.
Using Eqs.~(\ref{eq:Nsig}), (\ref{eq:NMS}), and (\ref{eq:hard})
in Eq.~(\ref{eq:Nmass}), we can obtain the same result as
that in GSS\cite{GASSER88}. In particular, the well-known
nonanalytic contribution to the nucleon mass is the leading term in
\begin{equation}
       \Sigma^{\overline{\rm MS}}_{\pi\rm N}(k)
           \Big |_{\rlap/{\mkern-1mu} k = M}
       = -     {3 g_{\rm A}^2 m_\pi^3 \over 32\pi f_\pi^2}
              \Big( 1 - {m_\pi \over2\pi M} + O(m_\pi^2) \Big) \ .
\end{equation}

Finally, let us briefly consider  the $\pi N$
scattering amplitude up to order $Q^3$. According to Eq.~(\ref{eq:chcnt}),
we need to calculate tree diagrams constructed from vertices
with $d_i+{1\over 2} n_i \leq 4$ and the soft parts of 
one-loop diagrams constructed from vertices with $d_i+{1\over 2} n_i =2$.
GSS have calculated the diagrams with
pions and nucleons\cite{GASSER88}, but their results are 
entangled with large contributions from anti-nucleons,
which must be absorbed into the parameters to obtain a useful
low-energy expansion. With the 
inclusion of the $\Delta$ isobar, the number of diagrams further
proliferate. The one-loop calculation in our approach is in progress.
Here we simply note that our calculation without including the nonanalytic
$O(Q^3)$ loop contributions has produced a good fit to the phase-shift 
data for energies up to the $\Delta$-resonance region\cite{ET96}, 
whereas the calculation
from heavy-baryon CHPT has not yet been satisfactory\cite{DATTA96}.

I thank P.J. Ellis,  R.J. Furnstahl, 
S. Jeon, J.I. Kapusta, and B.D. Serot for
stimulating discussions and useful comments. 
This work was supported by
the Department of Energy under grant No. DE-FG02-87ER40328.


\begin{thebibliography}{999}
%
\bibitem{WEINBERG79} S. Weinberg,
                     Physica {\bf A96}, 327 (1979).
%
\bibitem{GASSER84} J. Gasser and H. Leutwyler,
                   Ann.\ Phys. {\bf 158}, 142 (1984);
                   Nucl.\ Phys.\ {\bf B250}, 465,517,539 (1985).
%
\bibitem{WEINBERG90}S. Weinberg,
                Phys.\  Lett.\ {\bf B251}, 288 (1990);
                  Nucl.\ Phys.\ {\bf B363}, 3 (1991).
%
\bibitem{GASSER88} J. Gasser, M. E. Sainio, and A. Svarc,
             Nucl.\ Phys.\ {\bf B307}, 779 (1988).
%
\bibitem{JENKINS91} E. Jenkins and A. V. Manohar,
                Phys.\  Lett.\ {\bf B255}, 558 (1991).
%
\bibitem{LUTZ} M. Lutz is also pursuing a relativistic approach;
               see hep-ph/9606301.
%
\bibitem{ECKER96} G. Ecker and M. Mojzis,
                Phys.\  Lett.\ {\bf B365}, 312 (1996).
%
\bibitem{SEROT86} B. D. Serot and J. D. Walecka,
                  Adv.\ Nucl.\ Phys.\ {\bf 16}, 1 (1986);
         B. D. Serot, Rep.\ Prog.\ Phys.\ {\bf 55}, 1855 (1992).
%
\bibitem{FST95} R. J. Furnstahl, H.-B. Tang, and B. D. Serot,
                 Phys.\ Rev.\ {\bf C52}, 1368 (1995).
%
\bibitem{GEORGI91} H. Georgi,
              Nucl. Phys. {\bf B361}, 339 (1991).
%
\bibitem{ARZT95} C. Arzt, 
                Phys.\  Lett.\ {\bf B342}, 189 (1995).
%
\bibitem{MANOHAR84} A. V. Manohar and H. Georgi,
          Nucl.\ Phys.\ {\bf B234}, 189 (1984).
%
\bibitem{GEORGI93b} H. Georgi,
               Phys. Lett. {\bf B298}, 187 (1993).
%
\bibitem{TE96} H.-B. Tang and P. J. Ellis, hep-ph/9606432,
                  Phys.\  Lett.\ {\bf B}, in press.
%
\bibitem{ET96}  P. J. Ellis and H.-B. Tang,
          ``Pion--Nucleon Scattering at Low Energies'',
          hep-ph/9609459.
%
\bibitem{LEPAGE89} G. P. Lepage,
% ``What is Renormalization?'', 
        in {\it From Actions to Answers} (TASI-89), edited by
        T. DeGrand and D. Toussaint 
         (World Scientific, Singapore, 1989), p.483.
%
\bibitem{COLLINS84} J. C. Collins, 
              {\it Renormalization} 
        (Cambridge University Press, Cambridge, 1984).
%
\bibitem{BERNARD92} V. Bernard, N. Kaiser, J. Kambor,
          and U. G. Meissner, Nucl. Phys. {\bf B388}, 315 (1992).
%
\bibitem{DATTA96} A. Datta and S. Pakvasa, hep-ph/9606277.
%
\end{thebibliography}
\end{document}